\documentclass[twocolumn,aps,pra,longbibliography,superscriptaddress,nofootinbib,10pt]{revtex4-1}
\usepackage[latin1]{inputenc}
\usepackage{graphicx,dcolumn,bm}
\usepackage{subfigure}
\usepackage{multirow}
\usepackage{array}
\usepackage{arydshln}
\usepackage{color}
\usepackage[colorlinks,linkcolor=blue,citecolor=blue,hyperindex]{hyperref}
\usepackage{amsfonts}
\usepackage{amssymb}
\usepackage{amsmath}
\usepackage{latexsym}

\usepackage{simplewick}
\usepackage{epstopdf}
\usepackage{multirow}
\usepackage{float}
\usepackage[table]{xcolor}
\usepackage{multibib}
\usepackage{mathrsfs}
\usepackage[sans]{dsfont}
\usepackage[mathscr]{eucal}

\usepackage[normalem]{ulem}
\usepackage{epsfig}
\definecolor{Blue}{rgb}{0.0,0.0,1}
\definecolor{Red}{rgb}{1,0.0,0.0}
\definecolor{Green}{rgb}{0,0.5,0.0}
\usepackage{tikz}
\usetikzlibrary{decorations.pathmorphing}
\usetikzlibrary{arrows}
\usetikzlibrary{intersections,shapes.arrows}
\usetikzlibrary{calc}
\usetikzlibrary{quotes,angles}
\usepackage{nicefrac}
\usepackage{pgfplots}
\usepgfplotslibrary{fillbetween}
\pgfplotsset{compat=1.13,colormap={violetnew}{rgb=(0.293416, 0.0574044, 0.529412) rgb=(0.394818,0.233715,0.671945) rgb =(0.49622,0.410025,0.814477) rgb=(0.588672,0.567494,0.910066) rgb=(0.663226,0.687282,0.911765) rgb=(0.73778,0.807069,0.913465) rgb=(0.807267,0.861883,0.894034) rgb=(0.874222,0.884211,0.864039) rgb=(0.941176, 0.906538, 0.834043)}}
\usepgfplotslibrary{groupplots} 
\usepgfplotslibrary[groupplots] 
\usetikzlibrary{pgfplots.groupplots} 
\usetikzlibrary[pgfplots.groupplots] 
\usepgfplotslibrary{statistics}
\usepackage{pgfplotstable}
\tikzset{jumpdot/.style={mark=*,solid},excl/.append style={jumpdot,fill=white},incl/.append style={jumpdot,fill=black}}
\begin{document}

\title{Fidelity-based distance bounds for $N$-qubit approximate quantum error correction}
\author{Guilherme Fiusa}
\affiliation{Instituto de F\'{i}sica de S\~{a}o  Carlos, Universidade de S\~{a}o Paulo, CP 369, 13560-970, S\~{a}o Carlos, S\~{a}o Paulo, Brazil}
\author{Diogo O. Soares-Pinto}
\affiliation{Instituto de F\'{i}sica de S\~{a}o  Carlos, Universidade de S\~{a}o Paulo, CP 369, 13560-970, S\~{a}o Carlos, S\~{a}o Paulo, Brazil}
\author{Diego Paiva Pires}
\affiliation{Departamento de F\'{i}sica, Universidade Federal do Maranh\~{a}o, Campus Universit\'{a}rio do Bacanga, 65080-805, S\~{a}o Lu\'{i}s, Maranh\~{a}o, Brazil}

\begin{abstract}
The Eastin-Knill theorem is a central result of quantum error correction theory and states that a quantum code cannot correct errors exactly, possess continuous symmetries, and implement a universal set of gates transversely. As a way to circumvent this result, there are several approaches in which one gives up on either exact error correction or continuous symmetries. In this context, it is common to employ a complementary measure of fidelity as a way to quantify quantum state distinguishability and benchmark approximations in error correction. Despite having useful properties, evaluating fidelity measures stands as a challenging task for quantum states with a large number of entangled qubits. With that in mind, we address two distance measures based on the sub- and superfidelities as a way to bound error approximations, which in turn require a lower computational cost. We model the lack of exact error correction to be equivalent to the action of a single dephasing channel, evaluate the proposed fidelity-based distances both analytically and numerically, and obtain a closed-form expression for a general $N$-qubit quantum state. We illustrate our bounds with two paradigmatic examples, an $N$-qubit mixed GHZ state and an $N$-qubit mixed $W$ state.
\end{abstract}

\maketitle


\section{Introduction}
\label{sec:00000000001}

Quantum computers are among the most anticipated technological novelties of the present century. Their expected applicability ranges from the development of quantum algorithms to solve classically intractable physical problems~\cite{Shor1994,Grover1996}, to the efficient simulation of many-particle quantum systems~\cite{Reiher2017,Wecker2015}. However, a quantum computer with many qubits, i.e., from several dozens to a few hundred in the noisy intermediate-scale quantum (NISQ) era~\cite{Preskill2018nisq}, is subject to interaction with the environment in such a way that the computations are unreliable. It is necessary to employ a scheme that allows for reliable computations, providing fault tolerance~\cite{Shor1996,Aharonov1996,Kitaev2003fault}.

Quantum error correction (QEC) arises as a mechanism to accomplish such a task~\cite{Shor1995,Gottesman1996,PhysRevLett.77.198,Gottesman1997stabilizer,Knill1997,Steane1996,Terhal2015}. The idea consists of encoding information by using entanglement in a such way that the encoded information is protected against noise, yielding reliable computations. Furthermore, one is interested not only in maintaining reliable information but also in performing operations. Hence, despite seeking isolation with respect to the environment, a QEC scheme should provide ways to allow performing operations. In this regard, a convenient way to guarantee QEC while performing operations comes from transversal gates in quantum error-correcting codes~\cite{Yoshida2015,Bravyi2013}. Transversal gates act independently on each qubit in such a way that a faulty gate will only compromise single qubits, which means that errors are not spread out throughout the computation. 

A drawback from error-correcting codes with transversal gates comes from the no-go theorem derived by Eastin and Knill~\cite{EastinKnill2009}, which states that a quantum error-correcting code cannot exactly correct errors, possess continuous symmetries, and allow the implementation of a universal set of transversal gates. This result suggests that one must give up on either continuous symmetries of quantum error-correcting codes (covariant codes) or exact QEC. Motivated by the wide variety of phenomena in which covariant codes play a role, such as quantum reference frames~\cite{Woods202002,Hayden2021}, resource theories~\cite{Bartlett2007,Gour2008}, and quantum gravity~\cite{Pastawski2015,Almheiri2015,Harlow2021} via the anti-de Sitter (AdS)-conformal field theory (CFT) correspondence~\cite{Maldacena1999,Aharony2000}, we choose to give up on exact error correction. In other words, throughout the paper, we deal with covariant codes with transversal gates but that correct errors only approximately.

Lately, there has been a great deal of interest in quantifying the lack of exactness (i.e., the approximation) in quantum error correction. In particular, methods ranging from quantum metrology to quantum resource theories have been employed in this context~\cite{Zhou2021,Kubica2021,Faist2020,Zhou2020b,arXiv:2103.01876,Zhou2021b,Zhou2021c,arXiv:2206.11086,PhysRevA.56.2567,PhysRevA.81.062342,PhysRevLett.104.120501,PhysRevA.86.012335,PhysRevA.89.022316}. Although, in principle, any distinguishability measure could be employed as a figure of merit for the lack of exact QEC, fidelity has several good properties which make it an attractive choice. Overall, we refer to ``fidelity measures'' as an umbrella term that encompasses several measures that include, for example, Uhlmann's fidelity or even the so-called quantum infidelity~\cite{Nielsen_Chuang_infor_geom,Ingemar_Bengtsson_Zyczkowski}. Nonetheless, for a large number of entangled qubits, evaluating Uhlmann's fidelity is a challenging task, as it requires spectral properties of density matrices whose dimensions grow exponentially with the size of the system.

In this work, we address this issue by proposing two measures based on the sub- and superfidelities~\cite{Horodecki2008,Mendonca2008} to bound the approximation error in quantum error correction. The proposed measures establish lower and upper bounds to the typical fidelity error measure, while their evaluation requires a low computational cost. Since our measures are derived from the quantum fidelity, they inherit several useful properties. In particular, the measure based on the superfidelity defines a {\it bona fide} metric in the space of quantum states, and therefore can be employed to study underlying geometric concepts. We put our bounds to the test by modeling the approximation in error correction to be an effective dephasing channel. We evaluate the bounds for two paradigmatic examples, namely the $N$-qubit mixed GHZ state and the $N$-qubit mixed $W$ state. We also provide exact results for any quantum state considering our setup.

The text is organized as follows. In Sec.~\ref{sec:00000000002}, we discuss the main properties of covariant codes and approximate quantum error correction, laying down the fundamental ideas and the problem we address. In Sec.~\ref{sec:00000000003}, we review the main properties of fidelity measures, justifying their use and motivating our proposal. In Sec.~\ref{sec:00000000004}, we discuss our proposed bounds for approximate quantum error correction. In Sec.~\ref{sec:00000000005}, we consider the approximate error in quantum error correction to be modeled by a dephasing channel, and illustrate our proposal by evaluating the bounds for both the $N$-qubit mixed GHZ and $W$ states. To support our predictions, we provide analytical expressions and numerical simulations accordingly. In Sec.~\ref{sec:00000000006}, we summarize our results and present the concluding remarks.


\section{Covariant codes and approximate quantum error correction}
\label{sec:00000000002}

In this section, we briefly review the main concepts concerning the subject of approximate quantum error correction. A quantum code consists of a physical system $A$, a logical system $L$, and a completely positive and trace-preserving (CPTP) map called the encoding channel, which maps the logical system into the physical one. Both systems have a corresponding finite-dimensional Hilbert space, denoted by $\mathcal{H}_{A}$ (physical) and $\mathcal{H}_{L}$ (logical). We assume that the physical system consists of $N$ smaller subsystems such that $A = \otimes_{i=1}^{N}A_{i}$. We denote the encoding channel as $\mathcal{E}_{A \leftarrow L}$ and say that the code is covariant if~\cite{RevModPhys.91.025001}
\begin{equation}
\label{eq:00000000001}
\mathcal{E}_{A \leftarrow L} \circ \mathcal{U}_{L}^{\theta} = \mathcal{U}_{A}^{\theta} \circ  \mathcal{E}_{A \leftarrow L},
\end{equation}
where $\mathcal{U}_{L, A}^{\theta}$ implements the unitary symmetry transformation in the logical, physical subspace. The superscript index $\theta \in G$ denotes the group parameter.

Next, a covariant code is error correcting if, given a CPTP map $\mathcal{N}_{A}$ which acts upon the physical system $A$ and models the noise, there exists another CPTP map $\mathcal{R}_{L \leftarrow A}$ such that~\cite{Zhou2021,PRXQuantum.3.020314}
\begin{equation}
\label{eq:00000000002}
\mathcal{R}_{L \leftarrow A} \circ \mathcal{N}_{A} \circ \mathcal{E}_{A \leftarrow L} = \text{id}_{L},
\end{equation}
that is, for an initial state $\rho \in \mathcal{L}_{1}(\mathcal{H}_{L})$, with $\mathcal{L}_{1}(\mathcal{H}) = \{\rho \in \mathcal{H} \mid {\rho^{\dagger}} = \rho,~\rho\geq 0,~\text{Tr}(\rho) = 1\}$ being the convex set of Hermitian, positive semidefinite, and trace-one quantum states, the action of the quantum channels in the left-hand side of Eq.~\eqref{eq:00000000002} yield a final state $\rho$ equal to the initial one.  

In practice, the Eastin-Knill theorem forbids the existence of such codes and this is precisely the manifestation of approximate quantum error correction~\cite{EastinKnill2009}. Instead of recovering the identity channel of the logical subspace in Eq.~\eqref{eq:00000000002}, we have 
\begin{equation}
\label{eq:00000000003}
\mathcal{R}_{L \leftarrow A} \circ \mathcal{N}_{A} \circ \mathcal{E}_{A \leftarrow L} = \mathcal{I}_{L} \neq \text{id}_{L} ~.
\end{equation}
This means that the lack of exact error correction can be thought of as an effective channel $\mathcal{I}_{L}$ acting upon an arbitrary quantum state $\rho$~\cite{Zhou2021,Kubica2021,Faist2020}. In this scenario, the final recovered state will differ from the initial state. Hereafter, the effective quantum channel which models the approximate error correction is taken to be a global dephasing channel acting on an arbitrary number of qubits, each at its respective subspace.

In this context, the natural further step is to quantify how distinguishable the recovered quantum state is from the input state, in other words, to answer what is the difference between the effective quantum channel $\mathcal{I}_{L}$ and the logical identity channel $\text{id}_{L}$. Overall, the current literature employs fidelity measures as useful information-theoretic quantifiers to characterize the distinguishability between the initial $\rho$ and final $\mathcal{I}_{L}(\rho)$ quantum states~\cite{Zhou2021,Liu2020,Kubica2021,Faist2020}. In principle, any distinguishability measure should suffice; however, as we will see, fidelity measures have interesting properties which justify their use.


\section{Fidelities, sub- and superfidelities}
\label{sec:00000000003}

Fidelity is a measure of distinguishability between quantum states. As such, it is defined as~\cite{Jozsa1994,Schumacher1995,Uhlmann1976,Alberti1983b, Liang2019}
\begin{equation}
\label{eq:00000000004}
F(\rho, \sigma) = \left[\text{Tr}\left(\sqrt{\sqrt{\rho}\sigma \sqrt{\rho}} \right)\right]^{2} ~.
\end{equation}
As a way of distinguishing quantum states, fidelity measures have several desirable properties that motivate their use as a figure of merit in our context. In detail, for all quantum states $\{ {\rho_l} \}_{l = 1,\ldots,4} \in \mathcal{L}_{1}(\mathcal{H})$, it satisfies (i) positivity, $0 \leq F(\rho_{1}, \rho_{2}) \leq 1$; (ii) symmetry, $F(\rho_{1},\rho_{2}) = F(\rho_{2}, \rho_{1})$; (iii) unitary invariance, $F(\rho_{1}, \rho_{2}) = F(V\rho_{1} V^{\dagger}, V \rho_{2} V^{\dagger})$, for any unitary $V^{\dagger} = V^{-1}$; (iv) concavity, $F(\rho_{1}, \mu \rho_{2} + (1 - \mu) \rho_{3}) \geq \mu F(\rho_{1}, \rho_{2}) + (1 - \mu) F(\rho_{1}, \rho_{3})$, with $0 \leq \mu \leq 1$; (v) multiplicativity, $F(\rho_{1} \otimes \rho_{2}, \rho_{3} \otimes \rho_{4}) = F(\rho_{1}, \rho_{3})F(\rho_{2}, \rho_{4})$; and (vi) monotonicity under CPTP maps, $F(\rho, \sigma) \leq F(\mathcal{E}(\rho), \mathcal{E}(\sigma))$, $\forall \ \mathcal{E}(\bullet) \in \mathcal{L}(\mathcal{H})$, with $\mathcal{L}(\mathcal{H})$ denoting the set of linear bounded operators on $\mathcal{H}$.

In what we call the standard approach, the error related to the approximation in quantum error correction is defined in terms of the so-called Bures distance as follows~\cite{Ingemar_Bengtsson_Zyczkowski,Bures_135_199_1969}
\begin{equation}
\label{eq:00000000005}
\epsilon(\mathcal{I}_{L}, \text{id}_{L}) := \sqrt{1 - F(\mathcal{I}_{L}, \text{id}_{L})} ~,
\end{equation}
that is, one calculates the square root of the infidelity between the code $\mathcal{I}_{L}(\bullet)$ and the logical identity $\text{id}_{L}(\bullet)$ acting implicitly on a quantum state. If the code recovers precisely the logical identity, it means that the error correction is exact and thus the approximation error in Eq.~\eqref{eq:00000000005} is zero. On the other hand, if the code has orthogonal support concerning the logical identity, the fidelity is zero and the approximation error is maximum. 

We point out that, although fidelity stands as a useful distinguishability measure within the subject of quantum error correction, it exhibits drawbacks that hinder its applicability scope. The definition of fidelity in Eq.~\eqref{eq:00000000004} requires prior knowledge of the spectral properties of the input and output $N$-qubit states described by density matrices whose dimension scales exponentially with the system size. As a result, evaluating fidelity for higher dimensional systems is a challenging task, which means that the standard approach of defining the error approximation is of limited use when considering many qubits.

From now on, we make use of two information-theoretic quantifiers originally proposed in Refs.~\cite{Horodecki2008,Mendonca2008}, namely, the so-called subfidelity defined as
\begin{equation}
\label{eq:00000000006}
E(\rho,\sigma) := \text{Tr}(\rho \sigma) + \sqrt{2{[\text{Tr}(\rho \sigma)]^2} - 2\text{Tr}(\rho \sigma \rho \sigma)} ~,
\end{equation}
and also the superfidelity, given by
\begin{equation}
\label{eq:00000000007}
G(\rho,\sigma) := \text{Tr}(\rho \sigma) + \sqrt{[1-\text{Tr}(\rho^{2})][1-\text{Tr}(\sigma^{2})]} ~.
\end{equation}
In particular, the superfidelity for quantum states $\rho$, $\sigma$ is lower bounded in terms of the trace distance as $G(\rho,\sigma) \geq 1 - (1/2)\| {\rho} - {\sigma}\|_1$, with ${\| A \|_1} = \text{Tr}(\sqrt{{A^{\dagger}}A})$ being the Schatten 1-norm~\cite{PhysRevA.79.024302}. Furthermore, it has been proved that the sub- and superfidelities impose lower and upper bounds to the fidelity, respectively, written as~\cite{Horodecki2008,Mendonca2008}
\begin{equation}
\label{eq:00000000008}
E(\rho,\sigma) \leq F(\rho,\sigma) \leq G(\rho,\sigma) ~.
\end{equation}
In addition, for all quantum states $\{ {\rho_l} \}_{l = 1,\ldots,4} \in \mathcal{L}_{1}(\mathcal{H})$, it can be proved that the sub- and superfidelities exhibit the following properties: (i) positivity, $0 \leq E({\rho_1},{\rho_2}) \leq 1$ and $0 \leq G({\rho_1},{\rho_2}) \leq 1$; (ii) symmetry, $E({\rho_1},{\rho_2}) = E({\rho_2},{\rho_1})$ and $G({\rho_1},{\rho_2}) = G({\rho_2},{\rho_1})$; (iii) unitary invariance, $E({\rho_1},{\rho_2}) = E(V{\rho_1}{V^{\dagger}}, V{\rho_2}{V^{\dagger}})$ and $G({\rho_1}, {\rho_2}) = G(V{\rho_1}{V^{\dagger}}, V{\rho_2}{V^{\dagger}})$, for any unitary ${V^{\dagger}} = {V^{-1}}$; (iv) concavity, $E({\rho_1}, \mu {\rho_2} + (1 - \mu){\rho_3}) \geq  E({\rho_1},{\rho_2}) + (1 - \mu)E({\rho_1},{\rho_3})$, and $G({\rho_1}, \mu {\rho_2} + (1 - \mu){\rho_3}) \geq  G({\rho_1},{\rho_2}) + (1 - \mu)G({\rho_1},{\rho_3})$, with $0 \leq \mu \leq 1$; (v) subfidelity is submultiplicative, i.e., $E({\rho_1} \otimes{\rho_2}, {\rho_3}\otimes{\rho_4}) \leq E({\rho_1},{\rho_2})E({\rho_3},{\rho_4})$; and (vi) superfidelity is supermultiplicative, i.e., $G({\rho_1} \otimes{\rho_2}, {\rho_3}\otimes{\rho_4}) \geq G({\rho_1},{\rho_2})G({\rho_3},{\rho_4})$. Some remarks are now in order, and we are ready to address our proposal of fidelity-based distance measures useful to quantify approximations in quantum error correction.


\section{Bounds for approximate quantum error correction}
\label{sec:00000000004}

In the following, we consider the two fidelity-based distance measures related to the sub- and superfidelities, respectively,
\begin{equation}
\label{eq:000000000009}
\mathcal{D}_{\text{sub}}(\mathcal{I}_{L}, \text{id}_{L}):= \sqrt{1 - E(\mathcal{I}_{L}, \text{id}_{L})}
\end{equation}
and
\begin{equation}
\label{eq:000000000010}
\mathcal{D}_{\text{super}}(\mathcal{I}_{L}, \text{id}_{L}):= \sqrt{1 - G(\mathcal{I}_{L}, \text{id}_{L})} ~,
\end{equation}
where the quantum channels act upon some implicit quantum state and $\mathcal{I}_{L} = \mathcal{R}_{L \leftarrow A} \circ \mathcal{N}_{A} \circ \mathcal{E}_{A \leftarrow L}$. The relation between Eqs.~\eqref{eq:000000000009} and~\eqref{eq:000000000010} and the standard approach error approximation [see Eq.~\eqref{eq:00000000005}] comes from their definitions and the fidelity inequalities in Eq.~\eqref{eq:00000000008}, which yield
\begin{equation}
\label{eq:000000000011}
\mathcal{D}_{\text{super}}(\mathcal{I}_{L}, \text{id}_{L}) \leq \epsilon(\mathcal{I}_{L}, \text{id}_{L}) \leq \mathcal{D}_{\text{sub}}(\mathcal{I}_{L}, \text{id}_{L}) ~.
\end{equation}
Therefore, the distance measures based on the sub- and the superfidelities are upper and lower bounds, respectively. The inequalities are saturated when the sub- and the superfidelity recover the fidelity. This happens if at least one of the states is pure, or if one considers single-qubit states. In error-correction applications, none of these conditions is fully satisfied.

Interestingly, it has been shown that the superfidelity-based distance defines a genuine metric~\cite{Horodecki2008,Mendonca2008}. Indeed, for arbitrary quantum states $\rho_j \in \mathcal{L}_{1}(\mathcal{H})$, with $j = \{1,2,3\}$, it satisfies (i) semi-definite positiveness, $\mathcal{D}_{\text{super}}({\rho_1},{\rho_2}) \geq 0$ and $\mathcal{D}_{\text{super}}({\rho_1},{\rho_2}) = 0$ if and only if ${\rho_1} = {\rho_2}$; (ii) symmetry, $\mathcal{D}_{\text{super}}({\rho_1},{\rho_2}) = \mathcal{D}_{\text{super}}({\rho_2},{\rho_1})$; and (iii) triangle inequality,  $\mathcal{D}_{\text{super}}({\rho_1},{\rho_3}) \leq \mathcal{D}_{\text{super}}({\rho_1},{\rho_2}) + \mathcal{D}_{\text{super}}({\rho_2},{\rho_3})$~\cite{Mendonca2008}. The distance based on the subfidelity does not define a genuine metric because it is not positive semi-definite. More concretely, we put the bounds to the test by evaluating them while modeling the approximate error in quantum error correction as a dephasing channel acting on an $N$-qubit state.


\section{Application: Effective Dephasing Channel}
\label{sec:00000000005}

To illustrate the usefulness of the fidelity-based distance measures in Eqs.~\eqref{eq:000000000009} and~\eqref{eq:000000000010}, and also the chain of inequalities in Eq.~\eqref{eq:000000000011}, we model the lack of exact error correction as an effective quantum channel. To do so, we consider Eq.~\eqref{eq:00000000003} and set the quantum channel $\mathcal{I}_{L}(\bullet)$ to be a dephasing channel $\mathcal{E}_{\text{deph}}(\bullet)$. In addition, we consider the channel acting globally over a given $N$-qubit quantum state, which in turn can be written in terms of the Kraus representation as
\begin{align}
\label{eq:000000000012}
\mathcal{E}(\rho) &= {\sum_{{j_1},\ldots,{j_N}}} \, ({K_{j_1}}\otimes\ldots\otimes{K_{j_N}}) \rho ({K_{j_1}^{\dagger}}\otimes\ldots\otimes{K_{j_N}^{\dagger}}) ~,
\end{align}
for $j_{\ell} = \{0,1\}$, where $\ell = \{1,2,{\ldots},N\}$, with the Kraus operators $K_{0} = |0\rangle \langle 0| + \sqrt{1-p} |1\rangle \langle 1|$ and $K_{1} = \sqrt{p}|1\rangle \langle 1|$, while $0 \leq p \leq 1$ stands as the probability of noise being injected into the system. In the following, we address the issue of approximate error correction by employing the subfidelity and superfidelity distance measures by means of two paradigmatic probe states: the $N$-qubit GHZ mixed state, and the $N$-qubit $W$ mixed state. In the Appendix~\ref{sec:00000000007}, we present general, closed-form results for the sub- and superfidelities with respect to any given $N$-qubit quantum state that undergoes the action of the dephasing quantum operation. This can be accomplished by recasting the general $N$-qubit state in terms of the so-called Fano form~\cite{Ingemar_Bengtsson_Zyczkowski}, and exploiting algebraic properties of Pauli matrices.


\subsection{GHZ state}
\label{sec:00000000005A}

In the first example, we consider the initial mixed state to be
\begin{equation}
\label{eq:000000000013}
{\rho_{\text{GHZ}}} = \left(\frac{1 - \lambda}{2^N}\right)\mathbb{I} + \lambda\,|{\text{GHZ}_N}\rangle\langle{\text{GHZ}_N}| ~,
\end{equation}
with $0 \leq \lambda \leq 1$ being the mixing parameter, and $|{\text{GHZ}_N}\rangle$ denoting the GHZ state of $N$ particles defined as~\cite{Zeilinger1999}
\begin{equation}
\label{eq:000000000014}
|{\text{GHZ}_N}\rangle = \frac{1}{\sqrt{2}}\left(\,{|0\rangle^{\otimes N}} + {|1\rangle^{\otimes N}}\right) ~.
\end{equation}
The sub- and superfidelity distances for the states ${\rho_{\text{GHZ}}}$ and $\mathcal{E}({\rho_{\text{GHZ}}})$ are given by
\begin{equation}
\label{eq:000000000015}
{\mathcal{D}_{\text{sub}}} ({\rho_{\text{GHZ}}},\mathcal{E}({\rho_{\text{GHZ}}})) = \sqrt{1 - E({\rho_{\text{GHZ}}},\mathcal{E}({\rho_{\text{GHZ}}}))} ~,
\end{equation}
and
\begin{equation}
\label{eq:000000000016}
{\mathcal{D}_{\text{super}}} ({\rho_{\text{GHZ}}},\mathcal{E}({\rho_{\text{GHZ}}})) = \sqrt{1 - G({\rho_{\text{GHZ}}},\mathcal{E}({\rho_{\text{GHZ}}}))} ~,
\end{equation}
where the information-theoretic quantifiers $E(x,y)$ and $G(x,y)$ are given in Eqs.~\eqref{eq:00000000006} and~\eqref{eq:00000000007}, respectively. By considering the GHZ mixed state in Eq.~\eqref{eq:000000000013}, it can be verified that its purity becomes
\begin{equation}
\label{eq:000000000017}
\text{Tr}({\rho^2_{\text{GHZ}}}) = \frac{1}{2^N}\left(1 + ({2^N} - 1){\lambda^2}\right) ~,
\end{equation}
while the purity of the respective dephased state $\mathcal{E}({\rho_{\text{GHZ}}})$ [see Eq.~\eqref{eq:000000000012}] yields
\begin{align}
\label{eq:000000000018}
&\text{Tr}({\mathcal{E}({\rho_{\text{GHZ}}})^2}) = \nonumber\\
& = \frac{1}{2^N}\left(1 + ({2^{N - 1}} - 1 + {2^{N - 1}}{(1 - p)^N}){\lambda^2}\right) ~.
\end{align}
We point out that, for $p = 0$, Eq.~\eqref{eq:000000000018} recovers Eq.~\eqref{eq:000000000017} as a particular case. Next, the relative purity involving the density matrices ${\rho_{\text{GHZ}}}$ and $\mathcal{E}({\rho_{\text{GHZ}}})$ is written as follows:
\begin{align}
\label{eq:000000000019}
&\text{Tr}({\rho_{\text{GHZ}}} \mathcal{E}({\rho_{\text{GHZ}}})) = \nonumber\\
& = \frac{1}{2^N}\left(1 + ({2^{N - 1}} - 1 + {2^{N - 1}}{(1 - p)^{N/2}}){\lambda^2} \right) ~.
\end{align}
We emphasize that Eq.~\eqref{eq:000000000017} is also recovered from Eq.~\eqref{eq:000000000019} by setting the parameter $p = 0$, which means that the relative purity collapses into the purity in this limiting case. We find that Eqs.~\eqref{eq:000000000017},~\eqref{eq:000000000018}, and~\eqref{eq:000000000019} behave quadratically respective to the mixing parameter $\lambda$, while Eqs.~\eqref{eq:000000000018} and~\eqref{eq:000000000019} exhibit an $N$th-order polynomial dependence on the probability $p$. Finally, by performing lengthy calculations, one can verify the result
\begin{align}
\label{eq:000000000020}
&\text{Tr}({\rho_{\text{GHZ}}} \mathcal{E}({\rho_{\text{GHZ}}}) {\rho_{\text{GHZ}}} \mathcal{E}({\rho_{\text{GHZ}}})) = \nonumber\\
&= \frac{1}{2^{3N}} \left\{ 1 + {2^{N - 1}}\left({(2 + {(1 - p)^{N/2}})^2} + 3\,(1 - {2^{2 - N}})\right){\lambda^2} \right. \nonumber\\
&\left. + {2^N}({2^{N - 1}} - 1)\left( {( 2 + {(1 - p)^{N/2}} )^2} - 1 - {2^{3 - N}} \right){\lambda^3} \right.\nonumber\\
&\left. + {2^{N - 1}} [ {2^{2N - 2}}{(1 - p)^N} + {({2^{N - 1}} - 1)^2}{(2 + {(1 - p)^{N/2}})^2} \right.\nonumber\\
&\left.- (1 - {2^{1 - N}})({2^{2N - 1}} - 3) ]{\lambda^4} \right\} ~.
\end{align}

\begin{figure}[!t]
\begin{center}
\includegraphics[scale=0.925]{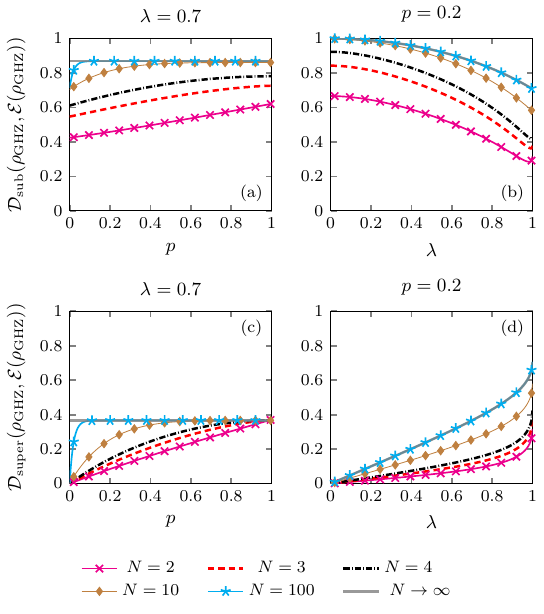}
\caption{(Color online) Plot of the sub- and superfidelity based distance measures for the mixed $N$-qubit GHZ state ${\rho_{\text{GHZ}}}$ [see Eq.~\eqref{eq:000000000013}] and the respective dephased density matrix $\mathcal{E}(\rho_{\text{GHZ}})$, for several system sizes $N$. [(a), (c)] Plots of the quantities ${\mathcal{D}_{\text{sub}, \text{super}}} ({\rho_{\text{GHZ}}},\mathcal{E}({\rho_{\text{GHZ}}}))$ as a function of the pro\-ba\-bi\-lity $0 \leq p \leq 1$, and choosing the mixing parameter value $\lambda = 0.7$. [(b), (d)] Fidelity-based distances ${\mathcal{D}_{\text{sub}, \text{super}}} ({\rho_{\text{GHZ}}},\mathcal{E}({\rho_{\text{GHZ}}}))$ as a function of the mixing parameter $0 \leq \lambda \leq 1$, for a fixed value $p = 0.2$.}
\label{fig:FIG01}
\end{center}
\end{figure}
%
In the following, we numerically investigate the be\-ha\-vior of both fidelity-based distance measures in Eqs.~\eqref{eq:000000000015} and~\eqref{eq:000000000016} by using the aforementioned analytical results in Eqs.~\eqref{eq:000000000017}--\eqref{eq:000000000020}. Figure~\ref{fig:FIG01} shows the plot of the distances based on subfidelity [see Eqs.~\eqref{eq:000000000015},~\eqref{eq:000000000019}, and~\eqref{eq:000000000020}] and superfidelity [see Eqs.~\eqref{eq:000000000016}--\eqref{eq:000000000019}] respective to the initial $N$-qubit mixed GHZ state, for different system sizes $N$. In Figs.~\ref{fig:FIG01}(a) and~\ref{fig:FIG01}(c), by setting the parameter $\lambda = 0.7$, one varies the parameter $0 \leq p \leq 1$. We find that the bounds increase as $p$ increases, i.e., the error approximation gets higher as we take larger values of $p$. This is precisely what one would expect from a practical point of view, as the more likely errors are to occur, the more likely the final and initial states will be distinct, which implies a larger error approximation (i.e., smaller sub- and superfidelities). In Figs.~\ref{fig:FIG01}(b) and~\ref{fig:FIG01}(d), one sets the parameter $p = 0.2$ and varies the mixing parameter $0 \leq \lambda \leq 1$. As $\lambda$ increases, the bounds become tighter; this behavior is expected because as the purity of the state increases, the sub- and superfidelities get closer to the fidelity, which in turn makes the upper and lower bounds to the error approximation closer.
\begin{figure}[!t]
\begin{center}
\includegraphics[scale=1.05]{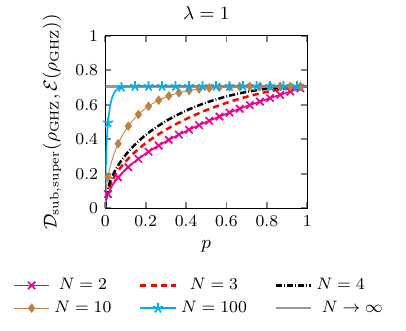}
\caption{(Color online) Plot of the sub- and superfidelity-based distance measures for the pure $N$-qubit GHZ state $\rho_{\text{GHZ}}$ [see Eq.~\eqref{eq:000000000014}] and the respective dephased density matrix $\mathcal{E}(\rho_{\text{GHZ}})$ as a function of the probability $p$. Note that both the distance measures approach the limiting value $1/\sqrt{2}$ for $N \rightarrow \infty$.}
\label{fig:FIG02}
\end{center}
\end{figure}

Next, Fig.~\ref{fig:FIG02} illustrates the behavior of both the sub- and superfidelity distances for the case of a probe $N$-qubit pure GHZ state, as a function of the parameter $0 \leq p \leq 1$. In this regard, for $\lambda = 1$, Eqs.~\eqref{eq:000000000017}--\eqref{eq:000000000020} imply that both the fidelity-based distance measures saturate to the Uhlmann fidelity as follows:
\begin{equation}
\label{eq:000000000021}
{\mathcal{D}_{\text{sub}, \text{super}}} ({\rho_{\text{GHZ}}},\mathcal{E}({\rho_{\text{GHZ}}})) = \sqrt{\frac{1 - {(1 - p)^{N/2}}}{2}} ~.
\end{equation}
Note that the range of possible values for the error approximation is influenced by the value of the probability. In this sense, smaller $p$ allows for a range of smaller values of the fidelity-based distance measures. However, for $p \rightarrow 1$, the allowed values increase, as it is more likely that errors occur, resulting in overall higher approximations. In particular, Eq.~\eqref{eq:000000000021} implies that $\mathcal{D}_{\text{sub}, \text{super}}({\rho_{\text{GHZ}}},\mathcal{E}({\rho_{\text{GHZ}}})) = {1}/{\sqrt{2}}$ for $N \rightarrow \infty$, regardless of the probability $0 \leq p \leq 1$. In Fig.~\ref{fig:FIG02}, this asymptotic behavior is observed by setting larger system sizes $N$.

\begin{figure}[!t]
\begin{center}
\includegraphics[scale=1.05]{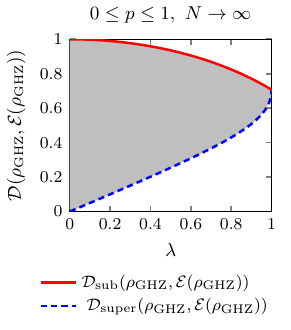}
\caption{(Color online) Plot of the sub- and superfidelity-based distance measures for the mixed $N$-qubit GHZ state $\rho_{\text{GHZ}}$ [see Eq.~\eqref{eq:000000000013}] and the respective dephased density matrix $\mathcal{E}({\rho_{\text{GHZ}}})$ as a function of the mixing parameter $\lambda$. Here we consider the case $N \rightarrow \infty$.}
\label{fig:FIG03}
\end{center}
\end{figure}

In Fig.~\ref{fig:FIG03}, we present the plots of both fidelity-based distance measures for the case of a multiparticle mixed GHZ state, as a function of the parameter $0 \leq \lambda \leq 1$, for $N \rightarrow \infty$. In this case, with the help of Eqs.~\eqref{eq:000000000017}--\eqref{eq:000000000020}, it can be verified that the subfidelity in Eq.~\eqref{eq:000000000015} approaches the asymptotic value
\begin{equation}
\label{eq:000000000022}
{\lim_{N\rightarrow\infty}}{\mathcal{D}_{\text{sub}}} ({\rho_{\text{GHZ}}},\mathcal{E}({\rho_{\text{GHZ}}})) = \sqrt{1 - \frac{\lambda^2}{2}} ~,
\end{equation}
while the superfidelity in Eq.~\eqref{eq:000000000016} becomes
\begin{align}
\label{eq:000000000023}
&{\lim_{N\rightarrow\infty}}{\mathcal{D}_{\text{super}}} ({\rho_{\text{GHZ}}},\mathcal{E}({\rho_{\text{GHZ}}})) =\nonumber\\
& = \sqrt{1 - \frac{\lambda^2}{2} - \sqrt{\frac{(1 - {\lambda^2})(2 - {\lambda^2})}{2}}} ~.
\end{align}
We point out that, for $N \rightarrow \infty$, Eqs.~\eqref{eq:000000000022} and~\eqref{eq:000000000023} are completely independent of the probability $0 \leq p \leq 1$ related to the dephasing channel. Figure~\ref{fig:FIG03} shows that, for small values of $\lambda$, the initial state is close to being maximally mixed, implying that the bounds are not very restrictive. In addition, as $\lambda \rightarrow 1$, the initial state becomes purer, to the point where it becomes a completely pure GHZ state for $\lambda = 1$. As a consequence, the sub- and the superfidelity become the fidelity, and the two bounds converge to the same value, given by the Uhlmann fidelity.


\subsection{$W$ state}
\label{sec:00000000005B}
\begin{figure}[!t]
\begin{center}
\includegraphics[scale=0.925]{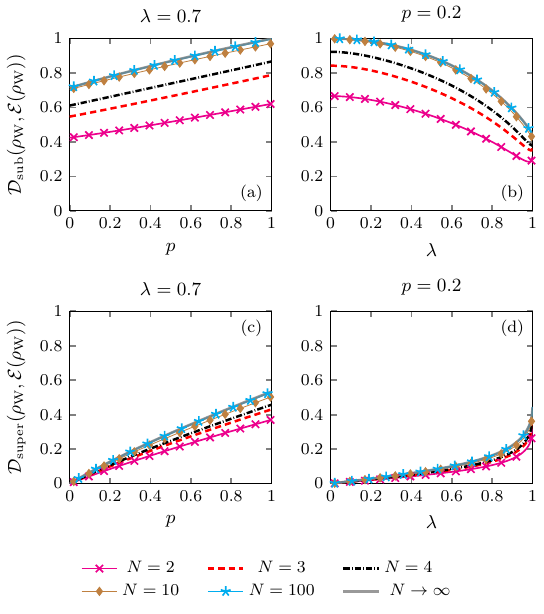}
\caption{(Color online) Plot of the sub- and superfidelity-based distance measures for the mixed $N$-qubit W state ${\rho_{\text{W}}}$ [see Eq.~\eqref{eq:000000000013}] and the respective dephased density matrix $\mathcal{E}({\rho_{\text{W}}})$, for several system sizes $N$. [(a), (c)] Plots of the quantities ${\mathcal{D}_{\text{sub}, \text{super}}} ({\rho_{\text{W}}},\mathcal{E}({\rho_{\text{W}}}))$ as a function of the pro\-ba\-bi\-lity $0 \leq p \leq 1$, and choosing the mixing parameter value $\lambda = 0.7$. [(b), (d)] Fidelity-based distances ${\mathcal{D}_{\text{sub}, \text{super}}} ({\rho_{\text{W}}},\mathcal{E}({\rho_{\text{W}}}))$ as a function of the mixing parameter $0 \leq \lambda \leq 1$, for a fixed value $p = 0.2$.}
\label{fig:FIG04}
\end{center}
\end{figure}

The second example we consider is the initial mixed state given by
\begin{equation}
\label{eq:000000000024}
{\rho_{\text{W}}} = \left(\frac{1 - \lambda}{2^N}\right)\mathbb{I} + \lambda\,|{W}\rangle\langle{W}| ~,
\end{equation}
with $0 \leq \lambda \leq 1$, and $|W\rangle$ denotes the $W$ state of $N$ particles given by~\cite{Cirac2000}
\begin{equation}
\label{eq:000000000025}
|{W}\rangle = \frac{1}{\sqrt{N}}\, {\sum_{l = 1}^N} ~ {{|0\rangle}^{\, \otimes{l - 1}}}\otimes{{|1\rangle}_l}\otimes{{|0\rangle}^{\, \otimes{N - l}}} ~.
\end{equation}
The fidelity-based distance measures given by the subfidelity and superfidelity for both states ${\rho_{\text{W}}}$ and $\mathcal{E}({\rho_{\text{W}}})$ are given as
\begin{equation}
\label{eq:000000000026}
{\mathcal{D}_{\text{sub}}} ({\rho_{\text{W}}},\mathcal{E}({\rho_{\text{W}}})) = \sqrt{1 - E({\rho_{\text{W}}},\mathcal{E}({\rho_{\text{W}}}))}
\end{equation}
and
\begin{equation}
\label{eq:000000000027}
{\mathcal{D}_{\text{super}}} ({\rho_{\text{W}}},\mathcal{E}({\rho_{\text{W}}})) = \sqrt{1 - G({\rho_{\text{W}}},\mathcal{E}({\rho_{\text{W}}}))} ~,
\end{equation}
with both the quantities $E(x,y)$ and $G(x,y)$ defined in Eqs.~\eqref{eq:00000000006} and~\eqref{eq:00000000007}, respectively. By considering the mixed $W$ state ${\rho_{\text{W}}}$ in Eq.~\eqref{eq:000000000024}, it can be verified that its purity becomes
\begin{equation}
\label{eq:000000000028}
\text{Tr}({\rho^2_{\text{W}}}) = \frac{1}{2^N}\left(1 + ({2^N} - 1){\lambda^2}\right) ~,
\end{equation}
while the purity of the respective dephased state $\mathcal{E}({\rho_{\text{W}}})$ yields
\begin{align}
\label{eq:000000000029}
&\text{Tr}({\mathcal{E}({\rho_{\text{W}}})^2}) = \nonumber\\
&= \frac{1}{2^N}\left[1 + \left({2^N} - 1 - \frac{{2^N}(N - 1)}{N}(2 - p)p\right){\lambda^2}\right] ~.
\end{align}
Note that, by setting $p = 0$, Eq.~\eqref{eq:000000000029} recovers Eq.~\eqref{eq:000000000028} as a particular case. Next, the relative purity involving the density matrices ${\rho_{\text{W}}}$ and $\mathcal{E}({\rho_{\text{W}}})$ is written as follows
\begin{align}
\label{eq:000000000030}
&\text{Tr}({\rho_{\text{W}}} \mathcal{E}({\rho_{\text{W}}})) = \nonumber\\
&=  \frac{1}{2^N}\left[1 + \left({2^N} - 1 - \frac{{2^N}(N - 1)}{N} p\right){\lambda^2}\right]  ~.
\end{align}
We emphasize that Eq.~\eqref{eq:000000000028} is also recovered from Eq.~\eqref{eq:000000000030} by setting the parameter $p = 0$, which means that the relative purity collapses into the quantum purity in this limiting case. We find that Eqs.~\eqref{eq:000000000028}--\eqref{eq:000000000030} behave quadratically respective to the mixing parameter $\lambda$. Finally, by performing lengthy calculations, one can verify the result
\begin{align}
\label{eq:000000000031}
&\text{Tr}({\rho_{\text{W}}} \mathcal{E}({\rho_{\text{W}}}) {\rho_{\text{W}}} \mathcal{E}({\rho_{\text{W}}})) = \nonumber\\
&= \frac{1}{{2^{3N}} \, {N^2}}\left\{ {N^2} + N [ {2^N}(6N - (N - 1)(6 - p)p) - 6N ]{\lambda^2} \right. \nonumber\\
&\left. + 2 [ 4{N^2} + {2^{2N}}(2N - (N - 1)p)(N - (N - 1)p) \right.\nonumber\\
&\left. - {2^N}N(6N - (N - 1)(6 - p)p) ]{\lambda^3} + [ {2^{3N}}{(N - (N - 1)p)^2} \right.\nonumber\\
&\left. - {2^{2N + 1}}(2N - (N - 1)p)(N - (N - 1)p) - 3{N^2} \right. \nonumber\\
&\left. + {2^N}N(6N - (N - 1)(6 - p)p) ]{\lambda^4}\right\} ~.
\end{align}

\begin{figure}[!t]
\begin{center}
\includegraphics[scale=1.05]{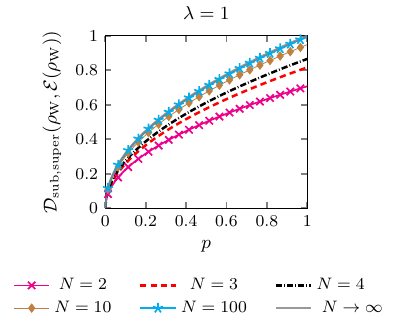}
\caption{(Color online) Plot of the fidelity-based distance measures ${\mathcal{D}_{\text{sub}, \text{super}}} ({\rho_{\text{W}}},\mathcal{E}({\rho_{\text{W}}}))$ [see Eq.~\eqref{eq:000000000032}] related to the $N$-qubit pure $W$ state $\rho_{\text{W}}$ and the respective dephased density matrix $\mathcal{E}({\rho_{\text{W}}})$, as a function of the probability $p$. In particular, for very large $N$, note that both distance measures approach the asymptotic value $\sqrt{p}$.}
\label{fig:FIG05}
\end{center}
\end{figure}

Next, by applying Eqs.~\eqref{eq:000000000028}--\eqref{eq:000000000031}, we present numerical simulations for the aforementioned fidelity-based distance measures. In Fig.~\ref{fig:FIG04}, we plot the quantifiers ${\mathcal{D}_{\text{sub}}} ({\rho_{\text{W}}},\mathcal{E}({\rho_{\text{W}}}))$ [see Eqs.~\eqref{eq:000000000026},~\eqref{eq:000000000030} and~\eqref{eq:000000000031}] and ${\mathcal{D}_{\text{super}}} ({\rho_{\text{W}}},\mathcal{E}({\rho_{\text{W}}}))$ [see Eqs.~\eqref{eq:000000000027},~\eqref{eq:000000000028},~\eqref{eq:000000000029} and~\eqref{eq:000000000030}] with respect to the initial mixed $N$-qubit $W$ state, for different system sizes. In Figs.~\ref{fig:FIG04}(a) and~\ref{fig:FIG04}(c), we set the parameter $\lambda = 0.7$, and consider the range $0 \leq p \leq 1$. We find that ${\mathcal{D}_{\text{sub,super}}}({\rho_{\text{W}}},\mathcal{E}({\rho_{\text{W}}}))$ increases as $p$ increases, with the superfidelity based distance approaching small values for small $p$. In Figs.~\ref{fig:FIG04}(b) and~\ref{fig:FIG04}(d), one sets the probability $p = 0.2$ while varying the mixing parameter $0 \leq \lambda \leq 1$. We find that the subfidelity-based (superfidelity-based) distance decreases (increases) as $\lambda$ increases. In other words, the bounds become tighter as expected when the purity of the initial state increases. On the one hand, for small values of $p$, we find that ${\mathcal{D}_{\text{sub}}} ({\rho_{\text{W}}},\mathcal{E}({\rho_{\text{W}}}))$ approaches the unity as one increases the system size $N$. On the other hand, note that ${\mathcal{D}_{\text{super}}} ({\rho_{\text{W}}},\mathcal{E}({\rho_{\text{W}}}))$ becomes zero as $p$ approaches zero.

Figure~\ref{fig:FIG05} shows the plot of the fidelity-based distance ${\mathcal{D}_{\text{sub}, \text{super}}} ({\rho_{\text{W}}},\mathcal{E}({\rho_{\text{W}}}))$ respective to the initial pure $W$ state of $N$ particles, as a function of the probability $0 \leq p \leq 1$, for different values of system sizes $N$. In detail, by setting $\lambda = 1$, one finds that both the distance-based subfidelity and superfidelity saturate to the fidelity-based distance measure, yielding
\begin{equation}
\label{eq:000000000032}
{\mathcal{D}_{\text{sub}, \text{super}}} ({\rho_{\text{W}}},\mathcal{E}({\rho_{\text{W}}})) = \sqrt{\frac{(N - 1)p}{N}} ~,
\end{equation}
which in turn holds for all $0 \leq p \leq 1$. On the one hand, we find that the fidelity-based distance measure vanishes for small values of $p$, regardless of the system size $N$ [see Fig.~\ref{fig:FIG05}]. On the other hand, for $p \rightarrow 1$, note that ${\mathcal{D}_{\text{sub}, \text{super}}} ({\rho_{\text{W}}},\mathcal{E}({\rho_{\text{W}}}))$ approaches unity for larger system sizes. We point out that, for the case $N \rightarrow \infty$, the fidelity-based distance in Eq.~\eqref{eq:000000000032} approaches the asymptotic value $\sqrt{p}$, with $0 \leq p \leq 1$. This is shown in Fig.~\ref{fig:FIG05}, where ${\mathcal{D}_{\text{sub}, \text{super}}} ({\rho_{\text{W}}},\mathcal{E}({\rho_{\text{W}}}))$ smoothly approaches the value $\sqrt{p}$ for larger $N$.

\begin{figure}[!t]
\begin{center}
\includegraphics[scale=0.95]{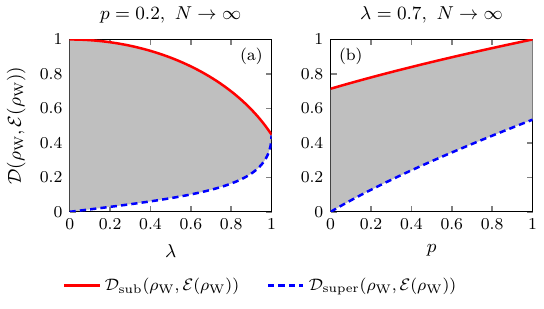}
\caption{(Color online) Plot of the sub- and superfidelity based distance measures for the $N$-qubit W state ${\rho_{\text{W}}}$ and the respective dephased density matrix $\mathcal{E}({\rho_{\text{W}}})$ [see Eqs.~\eqref{eq:000000000033} and~\eqref{eq:000000000034}]. Here we consider the case $N \rightarrow \infty$. We consider the asymptotic fidelity-based distance measures ${\mathcal{D}_{\text{sub}, \text{super}}} ({\rho_{\text{W}}},\mathcal{E}({\rho_{\text{W}}}))$ for values (a) $p = 0.2$ and $0 \leq \lambda \leq 1$, and (b) $\lambda = 0.7$ and $0 \leq p \leq 1$.}
\label{fig:FIG06}
\end{center}
\end{figure}

Next, Fig.~\ref{fig:FIG06} shows both subfidelity and superfidelity distances in the limiting case $N \rightarrow \infty$, for a given probe $N$-qubit mixed $W$ state with $0 \leq \lambda \leq 1$. We find that the distance measure based on subfidelity in Eq.~\eqref{eq:000000000026} approaches the asymptotic value 
\begin{equation}
\label{eq:000000000033}
{\lim_{N\rightarrow\infty}}{\mathcal{D}_{\text{sub}}} ({\rho_{\text{W}}},\mathcal{E}({\rho_{\text{W}}})) = \sqrt{1 - (1 - p){\lambda^2}} ~,
\end{equation}
for all $0 \leq p \leq 1$, while the superfidelity-based distance measure in Eq.~\eqref{eq:000000000027} becomes
\begin{align}
\label{eq:000000000034}
&{\lim_{N\rightarrow\infty}}{\mathcal{D}_{\text{super}}} ({\rho_{\text{W}}},\mathcal{E}({\rho_{\text{W}}})) = \nonumber\\ 
&= \sqrt{1 - (1 - p){\lambda^2} - \sqrt{(1 - {\lambda^2})(1 - {(1 - p)^2}{\lambda^2})}} ~.
\end{align}
In Fig.~\ref{fig:FIG06}(a), we set $p = 0.2$ and plot the fidelity-based distance measures in Eqs.~\eqref{eq:000000000033} and~\eqref{eq:000000000034} as a function of the mixing parameter $0 \leq \lambda \leq 1$. On the one hand, for small values of $\lambda$, we find that ${\mathcal{D}_{\text{super}}}({\rho_{\text{W}}},\mathcal{E}({\rho_{\text{W}}}))$ smoothly vanishes, while ${\mathcal{D}_{\text{sub}}}({\rho_{\text{W}}},\mathcal{E}({\rho_{\text{W}}}))$ approaches unity. On the other hand, as $\lambda \to 1$, the two quantities provide tighter bounds. In particular, for $\lambda = 1$, both the asymptotic quantities saturate to the standard fidelity measure. In Fig.~\ref{fig:FIG06}(b), with $\lambda = 0.7$, we show plots of the aforementioned asymptotic fidelity-based distance measures as a function of the probability $0 \leq p \leq 1$. Note that, for $p \rightarrow 1$, the subfidelity-based distance approaches unity. In Figs.~\ref{fig:FIG06}(a) and~\ref{fig:FIG06}(b), the shaded gray region sets the possible values of the standard fidelity-based distance measure, and thus the inequality in Eq.~\eqref{eq:000000000011} is fulfilled for all $0 \leq \lambda \leq 1$ and $0 \leq p \leq 1$.

\section{Conclusions}
\label{sec:00000000006}

In this work, motivated by the hardships of evaluating the fidelity for general $N$-qubit states, we provide an approach to quantify the error approximation in quantum error correction. We propose two distance measures based on sub- and superfidelities, and discuss their usefulness to bound the error approximation. We also provide concrete tests of our bounds. 

By considering the approximation in quantum error correction [see Eq.~\eqref{eq:00000000003}] to be modeled as a dephasing channel, we evaluate our bounds for two paradigmatic quantum states, namely both the mixed $N$-qubit GHZ and $W$ quantum states. We provide analytical results for both the sub- and superfidelities, and also present numerical simulations to support our predictions. In addition, we also provide closed-form expressions for the fidelity-based distance measures for general initial $N$-qubit quantum states undergoing the action of the dephasing channel [see the Appendix~\ref{sec:00000000007}]. We see our bounds as an important step in characterizing approximate quantum error correction, mainly because of the fact that the evaluation of both the sub- and superfidelity requires lower computational cost in contrast with the typical approach based on Uhlmann's fidelity.

For the initial mixed GHZ state, we observe that the probability parameter $p$ modulates the possible numerical values. As $p$ increases, the bounds increase accordingly. This behavior is expected because as the noise injection increases, the distance between the initial and final states also increases. The bounds become tighter as the mixture parameter $\lambda$ increase; in particular, they converge to the same value for a completely pure initial state. This is a consistency check that our bounds satisfy, because the sub- and superfidelities recover the fidelity for pure states. As the number of qubits increases, we observe that the two bounds converge to limiting values that depend only on the initial mixture parameter.

Next, for the initial mixed $W$ state, the parameter $p$ also modulates the numerical values, with the bounds increasing accordingly. We find that the bounds become more stringent as the purity of the initial state increases. On the one hand, for initial pure $W$ states with $\lambda = 1$, the bounds saturate to a value that depends on the square root of $p$. On the other hand, for $N$-qubit mixed $W$ states with larger $N$, the bounds reach an asymptotic value that depends on both $0 \leq \lambda \leq 1$ and $0 \leq p \leq 1$, in contrast with the case of the mixed GHZ states where the bounds depend only on the mixing parameter.

The theory of quantum error correction is seeing rapid development, and the study of approximate error correction is an important aspect of it. For example, recent applications of approximate error correction address the interplay of metrological bounds and global symmetries in AdS-CFT. Hence, as the relation between those areas has been a fruitful one, it would be interesting to further study the usefulness of our bounds in those contexts. Furthermore, from the fact that the distance measure based on the superfidelity [see Eq.~\eqref{eq:000000000010}] defines a {\it bona fide} metric on the space of quantum states~\cite{Mendonca2008}, it would be interesting to further reinterpret our results by exploiting the interplay between the subjects of quantum error correction and information geometry. Noteworthy, recent studies addressed the link of complexity with efficiency for designing quantum error-correcting codes within the framework of information geometry~\cite{PhysRevA.97.032323}. Remarkably, one finds that the efficiency and the information geometric complexity are related to the so-called entropic speed for optimal paths connecting initial and final states for a given physical process~\cite{PhysRevE.97.042110,PhysRevE.101.022110,PhysRevE.105.034143}. In order to address our results within the viewpoint of information geometry, we expect to investigate the family of underlying contractive Riemannian metrics (in a metric spaces sense) related to the superfidelity distance measure. Indeed, the so-called Morozova-\v{C}encov-Petz theorem predicts an infinite family of Riemannian metrics equipping the manifold of quantum states~\cite{1991_JSovietMath_56_2648,1996_LinAlgApl_244_81,1996_JMathPhys_37_2662,Ingemar_Bengtsson_Zyczkowski}. Within this perspective, for example, one can address the link between complexity and efficiency of quantum correction protocols for different Riemannian metrics. Finally, one can investigate how the sub- and superfidelity measures would behave for other quantum channels, e.g., depolarizing or amplitude damping. We intend to explore these ideas in further work.


\begin{acknowledgments}
This work was supported by the Brazilian ministries MEC and MCTIC, and the Brazilian funding agencies CNPq, and Coordena\c{c}\~{a}o de Aperfei\c{c}oamento de Pessoal de N\'{i}vel Superior--Brasil (CAPES) (Finance Code 001). D. O. S. P. acknowledges the Brazilian funding agencies CNPq (Grant No. 307028/2019-4), FAPESP (Grant No. 2017/03727-0), and the Brazilian National Institute of Science and Technology of Quantum Information (INCT-IQ) Grant No. 465469/2014-0. D. P. P. acknowledges Funda\c{c}\~{a}o de Amparo \`{a} Pesquisa e ao Desenvolvimento Cient\'{i}fico e Tecnol\'{o}gico do Maranh\~{a}o (FAPEMA). G. F. acknowledges support from Coordena\c{c}\~{a}o de Aperfei\c{c}oamento de Pessoal de N\'{i}vel Superior--Brasil (CAPES) (Finance Code 001).
\end{acknowledgments}

\setcounter{equation}{0}
\setcounter{table}{0}
\setcounter{section}{0}
\numberwithin{equation}{section}
\makeatletter
\renewcommand{\thesection}{\Alph{section}} 
\renewcommand{\thesubsection}{\Alph{section}.\arabic{subsection}}
\def\@gobbleappendixname#1\csname thesubsection\endcsname{\Alph{section}.\arabic{subsection}}
\renewcommand{\theequation}{\Alph{section}\arabic{equation}}
\renewcommand{\thefigure}{\arabic{figure}}
\renewcommand{\bibnumfmt}[1]{[#1]}
\renewcommand{\citenumfont}[1]{#1}

\section*{Appendix}


\section{General case}
\label{sec:00000000007}

In this appendix, we provide general results for the subfidelity and superfidelity distance measures for a given $N$-qubit probe state. We begin with the complete set of noncommuting operators $\{ {I_0}, {I_+}, {I_-}, {I_z} \}$, where $I_0 = (1/\sqrt{2})\mathbb{I}$, $I_z = (1/\sqrt{2}){\sigma_z}$, and ${I_{\pm}} = (1/2)({\sigma_x} \pm i {\sigma_y})$. Note that the normalized operators $I_k$ and $I_l$ are orthogonal to each other respective to the Hilbert-Schmidt inner product, i.e., one gets that $\text{Tr}({I_k^{\dagger}}{I_l}) = {\delta_{k,l}}$, for all $k,l \in \{0,\pm,z\}$. Hence, for a given general $N$-qubit probe state $\rho$, one readily finds that
\begin{equation}
\label{eq:0000000000A1}
\rho = {\sum_{{j_1},\ldots,{j_N}}}\, {a_{{j_1},\ldots,{j_N}}} \, {I_{j_1}}\otimes\ldots\otimes{I_{j_N}} ~,
\end{equation}
where
\begin{equation}
\label{eq:0000000000A2}
{a_{{j_1},\ldots,{j_N}}} = \text{Tr}[\rho ({I_{j_1}}\otimes\ldots\otimes{I_{j_N}})] ~.
\end{equation}
We consider the global dephasing map
\begin{equation}
\label{eq:0000000000A3}
\mathcal{E}(\rho) = {\sum_{{j_1},\ldots,{j_N}}}\, {a_{{j_1},\ldots,{j_N}}} \, {\mathcal{E}_1}({I_{j_1}})\otimes\ldots\otimes{\mathcal{E}_N}({I_{j_N}}) ~,
\end{equation}
for $j_{\ell} = \{0,\pm, z\}$ and $\ell = \{1,2,\ldots,N\}$, with the operation sum representation ${\mathcal{E}_l}({I_{j_l}}) = {\sum_{s = 0,1}} {K_s} {I_{j_l}} {K_s^{\dagger}}$, and the Kraus operators $K_{0} = |0\rangle \langle 0| + \sqrt{1-p} |1\rangle \langle 1|$ and $K_{1} = \sqrt{p}|1\rangle \langle 1|$, where $0 \leq p \leq 1$. In this case, it can be proved that
\begin{align}
\label{eq:0000000000A4}
&{\mathcal{E}_l}({I_{j_l}}) = {\delta_{{j_l},0}}{I_0} + {\delta_{{j_l},z}}{I_z} \nonumber\\
&+ \sqrt{1 - p} \, ({\delta_{{j_l},+}}{I_+} + {\delta_{{j_l},-}}{I_-}) ~.
\end{align}
The purity of the quantum state in Eq.~\eqref{eq:0000000000A1} reads as
\begin{equation}
\label{eq:0000000000A5}
\text{Tr}({\rho^2}) = {\sum_{{j_1},\ldots,{j_N}}}\, {\sum_{{k_1},\ldots,{k_N}}}\, {a_{{j_1},\ldots,{j_N}}} {a_{{k_1},\ldots,{k_N}}} \, {\prod_{l = 1}^N} \, \text{Tr}({I_{j_l}} {I_{k_l}}) ~,
\end{equation}
with
\begin{align}
\label{eq:0000000000A6}
&\text{Tr}({I_{j_l}} {I_{k_l}}) = {\delta_{{j_l},0}}{\delta_{{k_l},0}} + {\delta_{{j_l},z}}{\delta_{{k_l},z}} \nonumber\\
&+ {\delta_{{j_l},+}}{\delta_{{k_l},-}} + {\delta_{{j_l},-}}{\delta_{{k_l},+}} ~.
\end{align}
Next, the purity of the dephased state in Eq.~\eqref{eq:0000000000A3} is given by
\begin{align}
\label{eq:0000000000A7}
&\text{Tr}({\mathcal{E}({\rho})^2}) =\nonumber\\
& {\sum_{{j_1},\ldots,{j_N}}}\, {\sum_{{k_1},\ldots,{k_N}}}\, {a_{{j_1},\ldots,{j_N}}} {a_{{k_1},\ldots,{k_N}}} \, {\prod_{l = 1}^N} \, \text{Tr}(\mathcal{E}({I_{j_l}}) \mathcal{E}({I_{k_l}})) ~,
\end{align}
with 
\begin{align}
\label{eq:0000000000A8}
&\text{Tr}(\mathcal{E}({I_{j_l}}) \mathcal{E}({I_{k_l}})) = {\delta_{{j_l},0}}{\delta_{{k_l},0}} + {\delta_{{j_l},z}}{\delta_{{k_l},z}} \nonumber\\
&+ (1 - p) \, ({\delta_{{j_l},+}}{\delta_{{k_l},-}} + {\delta_{{j_l},-}}{\delta_{{k_l},+}}) ~.
\end{align}
The relative purity between the probe state $\rho$ and the dephased state $\mathcal{E}(\rho)$ is written as 
\begin{align}
\label{eq:0000000000A9}
&\text{Tr}(\rho\mathcal{E}({\rho})) = \nonumber\\
&{\sum_{{j_1},\ldots,{j_N}}}\, {\sum_{{k_1},\ldots,{k_N}}}\, {a_{{j_1},\ldots,{j_N}}} {a_{{k_1},\ldots,{k_N}}} \, {\prod_{l = 1}^N} \, \text{Tr}({I_{j_l}} \mathcal{E}({I_{k_l}})) ~,
\end{align}
where 
\begin{align}
\label{eq:0000000000A10}
&\text{Tr}({I_{j_l}} \mathcal{E}({I_{k_l}})) = {\delta_{{j_l},0}}{\delta_{{k_l},0}} + {\delta_{{j_l},z}}{\delta_{{k_l},z}} \nonumber\\
&+ \sqrt{1 - p} \, ({\delta_{{j_l},+}}{\delta_{{k_l},-}} + {\delta_{{j_l},-}}{\delta_{{k_l},+}}) ~.
\end{align}
Lastly, we evaluate the quantity
\begin{align}
\label{eq:0000000000A11}
&\text{Tr}(\rho\mathcal{E}({\rho})\rho\mathcal{E}({\rho})) = {\sum_{{j_1},\ldots,{j_N}}}\, {\sum_{{k_1},\ldots,{k_N}}} {\sum_{{q_1},\ldots,{q_N}}}\, {\sum_{{r_1},\ldots,{r_N}}} \, {a_{{j_1},\ldots,{j_N}}} \nonumber\\
& \times {a_{{k_1},\ldots,{k_N}}}{a_{{q_1},\ldots,{q_N}}} {a_{{r_1},\ldots,{r_N}}} \, {\prod_{l = 1}^N} \, \text{Tr}({I_{j_l}} \mathcal{E}({I_{k_l}}) {I_{q_l}} \mathcal{E}({I_{r_l}})) ~,
\end{align}
with
\begin{widetext}
\begin{align}
\label{eq:0000000000A12}
& \text{Tr}({I_{j_l}} \mathcal{E}({I_{k_l}}) {I_{q_l}} \mathcal{E}({I_{r_l}})) = \frac{1}{2} {\delta_{{j_l},0}}\left[ {\delta_{{k_l},0}}\left({\delta_{{q_l},0}}{\delta_{{r_l},0}} + \sqrt{1 - p} \, ( {\delta_{{q_l},+}}{\delta_{{r_l},-}} + {\delta_{{q_l},-}}{\delta_{{r_l},+}} ) + {\delta_{{q_l},z}}{\delta_{{r_l},z}}\right) \right. \nonumber\\
&\left. + \sqrt{1 - p} \, {\delta_{{k_l},+}}\left( \sqrt{1 - p} \, ( {\delta_{{q_l},0}} - {\delta_{{q_l},z}}){\delta_{{r_l},-}} + {\delta_{{q_l},-}}({\delta_{{r_l},0}} + {\delta_{{r_l},z}}) \right) \right. \nonumber\\
&\left. + \sqrt{1 - p} \, {\delta_{{k_l},-}}\left(\sqrt{1 - p} \, ( {\delta_{{q_l},0}} + {\delta_{{q_l},z}} ) {\delta_{{r_l},+}} + {\delta_{{q_l},+}} ({\delta_{{r_l},0}} - {\delta_{{r_l},z}} ) \right) \right. \nonumber\\
&\left. + {\delta_{{k_l},z}}\left({\delta_{{q_l},0}}{\delta_{{r_l},z}} + \sqrt{1 - p} \, ( {\delta_{{q_l},+}}{\delta_{{r_l},-}} - {\delta_{{q_l},-}}{\delta_{{r_l},+}} ) + {\delta_{{q_l},z}}{\delta_{{r_l},0}}\right) \right] \nonumber\\
& + \frac{1}{2} {\delta_{{j_l},+}}\left[ {\delta_{{k_l},0}}\left(\sqrt{1 - p} \, ( {\delta_{{q_l},0}} - {\delta_{{q_l},z}}){\delta_{{r_l},-}} + {\delta_{{q_l},-}}({\delta_{{r_l},0}} + {\delta_{{r_l},z}}) \right) \right. \nonumber\\
&\left. + \sqrt{1 - p} \, {\delta_{{k_l},-}}\left( ({\delta_{{q_l},0}} + {\delta_{{q_l},z}})({\delta_{{r_l},0}} + {\delta_{{r_l},z}}) + 2\sqrt{1 - p} \, {\delta_{{q_l},+}}{\delta_{{r_l},-}} \right) \right. \nonumber\\
&\left. - {\delta_{{k_l},z}}\left(\sqrt{1 - p} \, ( {\delta_{{q_l},0}} - {\delta_{{q_l},z}}){\delta_{{r_l},-}} + {\delta_{{q_l},-}}({\delta_{{r_l},0}} + {\delta_{{r_l},z}}) \right) \right] \nonumber\\
& + \frac{1}{2} {\delta_{{j_l},-}}\left[ {\delta_{{k_l},0}}\left(\sqrt{1 - p} \, ( {\delta_{{q_l},0}} + {\delta_{{q_l},z}}){\delta_{{r_l},+}} + {\delta_{{q_l},+}}({\delta_{{r_l},0}} - {\delta_{{r_l},z}}) \right) \right. \nonumber\\
&\left. + \sqrt{1 - p} \, {\delta_{{k_l},+}}\left( ({\delta_{{q_l},0}} - {\delta_{{q_l},z}})({\delta_{{r_l},0}} - {\delta_{{r_l},z}}) + 2\sqrt{1 - p} \, {\delta_{{q_l},-}}{\delta_{{r_l},+}} \right) \right. \nonumber\\
&\left. + {\delta_{{k_l},z}}\left(\sqrt{1 - p} \, ( {\delta_{{q_l},0}} + {\delta_{{q_l},z}}){\delta_{{r_l},+}} + {\delta_{{q_l},+}}({\delta_{{r_l},0}} - {\delta_{{r_l},z}}) \right) \right] \nonumber\\
& + \frac{1}{2} {\delta_{{j_l},z}}\left[ {\delta_{{k_l},0}}\left({\delta_{{q_l},0}}{\delta_{{r_l},z}} + \sqrt{1 - p} \, ( {\delta_{{q_l},+}}{\delta_{{r_l},-}} - {\delta_{{q_l},-}}{\delta_{{r_l},+}} ) + {\delta_{{q_l},z}}{\delta_{{r_l},0}}\right) \right. \nonumber\\
&\left. + \sqrt{1 - p} \, {\delta_{{k_l},+}}\left(\sqrt{1 - p} \, ( {\delta_{{q_l},0}} - {\delta_{{q_l},z}}){\delta_{{r_l},-}} + {\delta_{{q_l},-}}({\delta_{{r_l},0}} + {\delta_{{r_l},z}}) \right) \right. \nonumber\\
&\left. -  \sqrt{1 - p} \, {\delta_{{k_l},-}}\left(\sqrt{1 - p} \, ( {\delta_{{q_l},0}} + {\delta_{{q_l},z}}){\delta_{{r_l},+}} + {\delta_{{q_l},+}}({\delta_{{r_l},0}} - {\delta_{{r_l},z}}) \right) \right. \nonumber\\
&\left. + {\delta_{{k_l},z}}\left({\delta_{{q_l},0}}{\delta_{{r_l},0}} + \sqrt{1 - p} \, ( {\delta_{{q_l},+}}{\delta_{{r_l},-}} + {\delta_{{q_l},-}}{\delta_{{r_l},+}} ) + {\delta_{{q_l},z}}{\delta_{{r_l},z}}\right) \right]  ~.
\end{align}
\end{widetext}

Finally, by collecting the results in Eqs.~\eqref{eq:0000000000A5},~\eqref{eq:0000000000A7},~\eqref{eq:0000000000A9}, and~\eqref{eq:0000000000A11}, one finds analytical expressions for both the subfidelity and superfidelity distance measures for a general initial $N$-qubit probe state undergoing the action of the dephasing channel. We emphasize that these results solely depend on prior knowledge of the set of coefficients ${a_{{j_1},\ldots,{j_N}}}$ [see Eq.~\eqref{eq:0000000000A2}], which in turn can be readily obtained by writing the state $\rho$ in terms of the basis of operators $\{ {I_0}, {I_+}, {I_-}, {I_z} \}$.

We remind that the superfidelity distance measure ${\mathcal{D}_{\text{super}}}(\rho,\mathcal{E}(\rho)) = \sqrt{1 - G(\rho,\mathcal{E}(\rho))}$, with $G(\rho,\mathcal{E}(\rho)) = \text{Tr}(\rho\mathcal{E}(\rho)) + \sqrt{[1 - \text{Tr}({\rho^2})][1 - \text{Tr}({\mathcal{E}(\rho)^2})]}$, depends on the purities $\text{Tr}({\rho^2})$ [see Eqs.~\eqref{eq:0000000000A5} and~\eqref{eq:0000000000A6}] and $\text{Tr}({\mathcal{E}(\rho)^2})$ [see Eqs.~\eqref{eq:0000000000A7} and~\eqref{eq:0000000000A8}] of the probe state and the dephased state, respectively, and also the relative purity $\text{Tr}(\rho\mathcal{E}(\rho))$ [see Eqs.~\eqref{eq:0000000000A9} and~\eqref{eq:0000000000A10}] between these two quantum states. In turn, the subfidelity distance measure ${\mathcal{D}_{\text{sub}}}(\rho,\mathcal{E}(\rho)) = \sqrt{1 - E(\rho,\mathcal{E}(\rho))}$ stands as a function of the relative purity, with $E(\rho,\mathcal{E}(\rho)) = \text{Tr}(\rho\mathcal{E}(\rho)) + \sqrt{2 {[\text{Tr}(\rho\mathcal{E}(\rho))]^2} - 2 \text{Tr}(\rho\mathcal{E}(\rho)\rho\mathcal{E}(\rho))}$, but also depends on the quantity $\text{Tr}(\rho\mathcal{E}(\rho)\rho\mathcal{E}(\rho))$ [see Eqs.~\eqref{eq:0000000000A11} and~\eqref{eq:0000000000A12}].

In order to further investigate the quantity $\text{Tr}(\rho\mathcal{E}(\rho)\rho\mathcal{E}(\rho))$, we consider the spectral decomposition of $\rho$ and $\mathcal{E}(\rho)$ for a given $d$-dimensional quantum system. Let $\rho = {\sum_{r = 1}^d}\, {p_r}|{\psi_r}\rangle\langle{\psi_r}|$ be the spectral decomposition of the probe state, with the eigenvalues $\{ p_r \}_{r = 1,\ldots,d}$ satisfying $0 \leq {p_r} \leq 1$ and ${\sum_{r = 1}^d} \, {p_r} = 1$, and $\{ |{\psi_r}\rangle \}_{r = 1,\ldots,d}$ is the set of eigenstates of $\rho$, with $\langle{\psi_r}|{\psi_s}\rangle = {\delta_{r,s}}$, and ${\sum_{r = 1}^d} |{\psi_r}\rangle\langle{\psi_r}| = \mathbb{I}$. In addition, let $\mathcal{E}(\rho) = {\sum_{k = 1}^d} \, {\chi_k}|{\phi_k}\rangle\langle{\phi_k}|$ be the spectral decomposition of the dephased state, with the eigenvalues $\{ \chi_k \}_{k = 1,\ldots,d}$ satisfying $0 \leq {\chi_k} \leq 1$ and ${\sum_{k = 1}^d}\, {\chi_k} = 1$, while $\{ |{\phi_k}\rangle \}_{k = 1,\ldots,d}$ stand for the set of eigenstates of $\mathcal{E}(\rho)$, such that $\langle{\phi_k}|{\phi_l}\rangle = {\delta_{k,l}}$, and ${\sum_{k = 1}^d}|{\phi_k}\rangle\langle{\phi_k}| = \mathbb{I}$. Thus, one gets the following result:
\begin{align}
\label{eq:0000000000A13}
&\text{Tr}(\rho\mathcal{E}(\rho)\rho\mathcal{E}(\rho)) = \nonumber\\
&= {\sum_{r,s = 1}^d} \, {\sum_{k,l = 1}^d} \, {p_r}{p_s}{\chi_k}{\chi_l} \, \langle{\psi_r}|{\phi_k}\rangle \, {\langle{\psi_r}|{\phi_l}\rangle^*} \, \langle{\psi_s}|{\phi_l}\rangle \, {\langle{\psi_s}|{\phi_k}\rangle^*} \nonumber\\
&= {\sum_{r,k = 1}^d} \, {p_r^2}{\chi_k^2} \, {|\langle{\psi_r}|{\phi_k}\rangle|^4} \nonumber\\
&+ {\sum_{r \neq s }} \, {\sum_{k \neq l}} \, {p_r}{p_s}{\chi_k}{\chi_l} \, \langle{\psi_r}|{\phi_k}\rangle \, {\langle{\psi_r}|{\phi_l}\rangle^*} \, \langle{\psi_s}|{\phi_l}\rangle \, {\langle{\psi_s}|{\phi_k}\rangle^*}  ~.
\end{align}
Interestingly, note that the square of the relative purity between the states $\rho$ and $\mathcal{E}(\rho)$ is written as follows:
\begin{align}
\label{eq:0000000000A14}
&{[\text{Tr}(\rho\mathcal{E}(\rho))]^2} = {\sum_{r,s = 1}^d} \, {\sum_{k,l = 1}^d} \, {p_r}{p_s}{\chi_k}{\chi_l} \, {|\langle{\psi_r}|{\phi_k}\rangle|^2} \, {|\langle{\psi_s}|{\phi_l}\rangle|^2} \nonumber\\
&= {\sum_{r,k = 1}^d} \, {p_r^2}{\chi_k^2} \, {|\langle{\psi_r}|{\phi_k}\rangle|^4} \nonumber\\
&+ {\sum_{r \neq s}^d} \, {\sum_{k \neq l}^d} \, {p_r}{p_s}{\chi_k}{\chi_l} \, {|\langle{\psi_r}|{\phi_k}\rangle|^2} \, {|\langle{\psi_s}|{\phi_l}\rangle|^2} ~.
\end{align}
We note that the quantity $\text{Tr}(\rho\mathcal{E}(\rho)\rho\mathcal{E}(\rho))$ in Eq.~\eqref{eq:0000000000A13} somewhat resembles the square of the relative purity in Eq.~\eqref{eq:0000000000A14}, but the former exhibits an intricate combination of overlaps between states $|\psi_m\rangle$ and $|\phi_n\rangle$, for all $m,n = \{1,\ldots,d\}$. In the particular case of orthogonal vectors $|\psi_m\rangle$ and $|\phi_n\rangle$, i.e., $\langle{\psi_m}|{\phi_n}\rangle = 0$ for all $m,n = \{1,\ldots,d\}$, one finds that both Eqs.~\eqref{eq:0000000000A13} and~\eqref{eq:0000000000A14} vanish. We may understand the quantity $\text{Tr}(\rho\mathcal{E}(\rho)\rho\mathcal{E}(\rho))$ as a distinguishability measure of the states $\rho$ and $\mathcal{E}(\rho)$, somewhat similar to the standard relative purity $\text{Tr}(\rho\mathcal{E}(\rho))$.

To conclude, we comment on the particular case in which $\rho = |\psi\rangle\langle\psi|$ is an $N$-qubit pure state, while $\mathcal{E}(\rho) = \mathcal{E}(|\psi\rangle\langle\psi|)$ stands for the dephased mixed state. In this setting, one gets that $\text{Tr}(\rho\mathcal{E}(\rho)\rho\mathcal{E}(\rho)) = {\langle\psi|\mathcal{E}(\rho)|\psi\rangle^2} = {[\text{Tr}(\rho\mathcal{E}(\rho))]^2}$, i.e., one finds that the quantity reduces to the square of the relative purity between states $\rho$ and $\mathcal{E}(\rho)$. For two maximally distinguishable (orthogonal) states $\rho$ and $\mathcal{E}(\rho)$, both quantities become zero.




%

\end{document}